\documentclass[conference, a4paper]{IEEEtran}
\IEEEoverridecommandlockouts                              
\usepackage{graphicx}
\usepackage{multirow}
\usepackage{amsmath,amssymb,latexsym}
\usepackage{rotating}
\usepackage{lettrine}
\usepackage{textcomp}
\usepackage{siunitx}
\usepackage{tikz}
\usepackage{subcaption}

\usepackage{hyperref}
\usepackage{tikz}

\usepackage{xcolor}

\usepackage{lipsum}
\usepackage[pscoord]{eso-pic}
\newcommand{\placetextbox}[3]{
  \setbox0=\hbox{#3}
  \AddToShipoutPictureFG*{
    \put(\LenToUnit{#1\paperwidth},\LenToUnit{#2\paperheight}){\vtop{{\null}\makebox[0pt][c]{#3}}}%
  }%
}%

\usepackage{xcolor}

\newcommand \TD {T_\mathrm{D}}

\newcommand \AtB  {{A$\rightarrow$B} }
\newcommand \BtA  {{B$\rightarrow$A} }
\newcommand \alt  {{A$\sim$B} }
\newcommand \best {{A$^\wedge$B} }

\title{Performance Evaluation of Parallel Wi-Fi Redundancy with Deferral Techniques

\thanks{This work was partially supported by the European Union under the Italian National Recovery and Resilience Plan (NRRP) of NextGenerationEU, partnership on ``Telecommunications of the Future'' (PE00000001 - program ``RESTART'').}
}

\author{
    \IEEEauthorblockN{
    Gianluca Cena, 
    Pietro Chiavassa, 
    Stefano Scanzio 
    }
    \IEEEauthorblockA{
        National Research Council of Italy (CNR--IEIIT), Italy.} 
    Email: gianluca.cena@cnr.it, pietrochiavassa@cnr.it, stefano.scanzio@cnr.it
}

\begin{document}
\placetextbox{0.5}{1}{This is the author's version of an article that has been published.}
\placetextbox{0.5}{0.985}{Changes were made to this version by the publisher prior to publication.}
\placetextbox{0.5}{0.97}{The final version of record is available at \href{https://doi.org/10.1109/ETFA65518.2025.11205557}{https://doi.org/10.1109/ETFA65518.2025.11205557}}%
\placetextbox{0.5}{0.05}{Copyright (c) 2025 IEEE. Personal use is permitted.}
\placetextbox{0.5}{0.035}{For any other purposes, permission must be obtained from the IEEE by emailing pubs-permissions@ieee.org.}%

\maketitle
\thispagestyle{empty}
\pagestyle{empty}

\begin{abstract}
Wireless communication is increasingly used in industrial environments, since it supports mobility of interconnected devices.
Among the transmission technologies operating in unlicensed bands available to this purpose, 
Wi-Fi is certainly one of the most interesting, because of its high performance and the relatively low deployment costs.
Unfortunately, its dependability is often deemed unsuitable for real-time control systems.

In this paper, the use of parallel redundancy is evaluated from a quantitative viewpoint, by considering a number of performance indices that are relevant for soft real-time applications.
Analysis is carried out on a large dataset acquired from a real setup, to provide realistic insights on the advantages this kind of approaches can provide.
As will be seen, deferred parallel redundancy provides clear advantages in terms of the worst-case transmission latency,
at limited costs concerning the amount of consumed spectrum.
Hence, it can be practically exploited every time a wireless connection is included in a control loop.
\end{abstract}


\section{Introduction}
The ability to interconnect devices over the air permits lowering wiring harness tangibly, improving flexibility while decreasing installation cost, and becomes essential when mobility must be supported.
For these reasons, wireless network technologies are increasingly adopted also in industrial environments, where they enable, e.g., autonomous mobile robots (AMRs) to communicate among themselves and with the plant backbone, and hence the integration of all their activities at the shop-floor and in warehouses.
When high-performance connections are needed whose throughput and latency must resemble wired networks, i.e., (industrial) Ethernet, two solutions are practically available: cellular networks (5G) and \mbox{Wi-Fi} (IEEE 802.11) \cite{IEEE80211-24}.
While the former supports mobility over vast regions (potentially everywhere), Wi-Fi is targeted at interconnecting devices spread over areas of limited size, especially indoors, as happens in many production plants.
Furthermore, Wi-Fi communication equipment is generally less expensive than 5G and offers direct compatibility with Ethernet, since they share the same addressing space (MAC-48) and have similar maximum transfer units (MTU).

Unfortunately, as with every wireless technology, Wi-Fi is unable to ensure the same dependability as cables.
This depends essentially on two aspects: first, its basic medium access control (MAC) mechanism, known as distributed coordination function (DCF), is contention-based. 
Although it has been proven over the years to offer reliable behavior, collisions are unavoidable and grow as the overall traffic increases.
In fact, several improvements brought to Wi-Fi over the past two decades (PCF, HCCA, trigger frames) were meant to decrease the likelihood and impact of collisions.
The second aspect is related to the very nature of the wireless spectrum, which is affected by a number of different phenomena~\cite{10736552}, including electromagnetic noise, path loss, multipath fading, and so on.
What is worse, due to Wi-Fi operating in unlicensed bands, the presence of legacy wireless stations (STAs) that do not obey the newer access rules, or even devices using different transmission technologies~\cite{9750069}, cannot be excluded.
This can void many of the benefits of enhanced access techniques.

To deal with disturbance, a retransmission mechanism has been included in DCF, which repeatedly performs transmission attempts (spaced by randomly selected backoffs, whose duration increases exponentially) until either an acknowledgment is received from the recipient or the retry limit is exceeded.
Although retransmissions noticeably lower the likelihood of packet losses in non-adverse operating conditions, they can significantly increase latency, to the point that an appreciable number of packets may exceed their intended deadlines.
This is a major issue in real-time systems.

All recent advances to Wi-Fi were primarily meant to improve throughput and predictability for multimedia traffic.
In the following, instead, we will focus on industrial applications.
On the one hand, timing requirements are typically tighter than multimedia, with transmission deadlines as low as a few tens of milliseconds, and even less.
On the other hand, the amount of traffic generated and sent on air by every single node is noticeably lower: operations of control applications are known in advance and it is not unrealistic to assume that STAs are always working below saturation conditions.
Nevertheless, temporary phenomena that affect the wireless spectrum still exist and can lead to unpredictability.
Therefore, mechanisms capable of lowering the probability that packets are lost or delayed excessively are sought.
Clearly, no guarantees can be provided that deadlines will never be exceeded.
However, it is possible to increase the probability that they are met.

Several efforts were made in the past few years to improve Wi-Fi dependability and timeliness in such scenarios \cite{10034532}, some of which consider Wi-Fi 6 \cite{CANDELL-2024} and the recently defined Wi-Fi 7 \cite{JETMIR-24}.
Interestingly, the next version of this technology, known as Wi-Fi 8, is specifically geared at ensuring ultra high reliability (UHR) \cite{Galati-2024}.

In this paper, some approaches that exploit redundancy to improve dependability in digital communication networks are presented.
Generally speaking, two types of redundancy can be devised.
The former includes the spanning tree protocol (STP) and its rapid version (RSTP), customarily implemented in Ethernet, which suitably disable links so that the network topology is turned into a tree where a single physical path exists between any pair of nodes.
These approaches are mainly aimed at improving dependability: if a link or a switch becomes faulty, disabled links are re-enabled in such a way that connectivity can be restored.
In modern equipment, this is done rather quickly (a few milliseconds to  seconds), which is unlikely to disrupt communication in office automation scenarios.
However, this intervention time may be excessive for applications related to factory automation.
In these cases, solutions such as the parallel redundancy protocol (PRP) \cite{iec_industrial_2010} and the frame replication and elimination for redundancy (FRER) \cite{FRER}, explicitly defined in the context of time-sensitive networking (TSN), were introduced.
Such solutions ensure zero-intervention time and are collectively known as seamless redundancy.
This is achieved by sending the same frame on two (or more) separate paths and by removing duplicate copies before they reach the intended recipient.

The use of seamless redundancy in Wi-Fi for achieving frequency diversity was first proposed in \cite{WFCS12-RENT}.
In that work, a communication system is described where every packet is sent on two distinct Wi-Fi links according to PRP.
While very effective, this initial approach has a severe drawback, since it consumes twice as many resources (wireless spectrum) as non-redundant links.
In \cite{cena_enhanced_2014}, a reactive duplication avoidance (RDA) mechanism was described that 
uses cross acknowledgments (XACK) to terminate all ongoing Wi-Fi transmissions related to any given packet
as soon as a relevant ACK frame arrives on any link.
This technique was later improved by adding a deferral technique, in such a way that transmissions on the two channels do not start at the same time \cite{2016-TII-WiRed}. 
Doing so proactively increases the likelihood that ongoing transmissions can be aborted early, hence saving airtime.
A thorough analysis of the effects of deferral is available in \cite{2019-TWC-wired}.
Unlike that work, which relies on the ability of Wi-Fi controllers to support XACKs, this paper only considers ordinary Wi-Fi adapters.
Therefore, all the mechanisms described below can be implemented on commercial hardware and only require a limited amount of modifications to drivers.

The paper is structured as follows: in Section~\ref{sec:red} seamless redundancy for Wi-Fi is briefly summarized, while in Section~\ref{sec:exp} the experimental setup for acquiring the dataset used for post-analysis is described.
Section~\ref{sec:def} describes deferral techniques to lower spectrum consumption, and Section~\ref{sec:alt} introduces two enhancements for dynamically selecting the channel on which the next transmission should be performed.
Conclusions are finally drawn in Section~\ref{sec:conc}, together with future work.

\section{Seamless Redundancy}
\label{sec:red}
Industrial real-time applications customarily demand for dependability and timeliness.
A simple way to improve both in Wi-Fi is to exploit seamless redundancy.
In practice, two (or more) wireless links, which are associations between a given STA and a given access point (AP), are set up and jointly managed, and every packet is sent on all links by performing distinct transmission requests.
From the point of view of data-link layer users (either IP or specific reduced-profile industrial application protocols), these links are seen as a single redundant link.
Practically, seamless redundancy can be managed using existing protocols, like PRP or FRER.
These protocols take care of replicating every packet on the transmitting side, by creating several identical copies, and eliminating duplicates on the receiving side, so that no more than one copy is delivered to the user.
To do so, additional information is added to every frame by the sender, e.g., the redundancy control trailer 
in PRP, which uniquely identifies both the packet (e.g., by means of a sequence number) and the link on which the copy was sent.
In this way, late copies can be unambiguously identified by the recipient and discarded.

Individual links differ for the channel they use, and, for this reason, in the following they will often be referred to simply as channels.
In theory, they can use different Wi-Fi versions (e.g., Wi-Fi 4 on the $\SI{2.4}{GHz}$ band and Wi-Fi 5 on the $\SI{5}{GHz}$ band) and different protocol configurations (e.g., RTS/CTS usage).
Channels shall not overlap, so that they do not interfere with each other.
Having links as uncorrelated as possible is a prerequisite for making redundancy effective.
In practice, channels must be spaced widely enough, so that they are unlikely to suffer from the same electromagnetic phenomena.
The best solution is to choose them in different bands ($2.4$, $5$, and $\SI{6}{GHz}$), even though this is not strictly necessary.

The above arrangement closely resembles multi-link operation (MLO) in Wi-Fi 7, where a multi-link device (MLD), either an AP or a STA, 
is made up of two or more affiliated STAs.
In that case, the upper part of the MLD is known as U-MAC (there is only one per MLD), while the lower part foresees a distinct L-MAC for every affiliated link.
Although MLO supports, in principle, seamless redundancy (by using a specific TID-to-link mapping), it was not explicitly conceived for industrial applications.
The solution described in this article can be implemented on both new Wi-Fi~7 adapters and previous-generation adapters by bringing some software modifications to their drivers.

\subsection{Benefits and Drawbacks of Seamless Redundancy}
Redundancy always improves dependability, which empirically coincides with the packet delivery ratio (PDR).
Under the realistic assumption that channels are uncorrelated, the packet loss ratio (PLR) on the redundant link, defined as $1-$PDR, equals the product of the PLRs on individual channels.
In fact, a packet is definitely lost only if transmission fails on every channel of the redundant link.

When seamless redundancy is exploited, the transmission latency for a packet equals the delay experienced by the fastest copy.
Although this does not provide any strict guarantees that deadlines will be met, timeliness is likely to improve tangibly.
To minimize latency, transmission requests for the different copies of any given packet must be issued at the same time, as soon as it is made available by the upper layer.
In turn, every transmission on a specific link consists of one or more frame transmission attempts, performed by the MAC layer according to the conventional access rules that concern confirmed transmission services in Wi-Fi.

On the downside, sending every packet twice doubles spectrum consumption, which is not acceptable in many real scenarios, where unlicensed bands are also used for other purposes (office automation, multimedia streaming, etc.).

\subsection{Displacing Transmissions on Links}
A simple remedy to lower spectrum consumption, at the cost of reduced performance, 
is to start transmissions on distinct channels at different times.
For the sake of simplicity, from now on we will only consider the case of redundant links operating on two channels, which we denote as A and B.
One of them is selected as the \textit{primary} channel (P): on it, the transmission request is invoked as soon as the packet becomes available, according to the usual Wi-Fi rules.
Transmission on the other channel, we term \textit{secondary} (S), is delayed by a deferral time $\TD$.
If the transmission on P terminates correctly before $\TD$, transmission on S is not started at all.
Otherwise, transmission is also started on S, which may or may not overtake P.
Although in this paper we do not assume the presence of any abortion mechanisms that properly
support XACKs,
exploiting them when available always improves performance, as fewer spectrum resources are wasted.

In the following, $\TD$ was set to a fixed value. 
However, its duration should be selected based on the packet airtime, e.g.,
\begin{align}
    \TD = \beta \cdot d_{\min}^\mathrm{P},
\end{align}
where $d_{\min}^\mathrm{P}$ is the minimum duration for the considered frame on P while $\beta>1$ is a multiplicative factor.
Importantly, $d_{\min}^\mathrm{P}$ can be computed at runtime given the packet size, expressed as the length in bytes of the payload, 
and the data rate, as determined by the modulation and coding scheme (MCS), both of which are known by the Wi-Fi driver and chipset.

Selecting for $\TD$ a value shorter than $d_{\min}^\mathrm{P}$ is completely pointless, as doing so never prevents transmission on the secondary channel (which means that there are no benefits about spectrum consumption compared to simultaneous redundancy).
The larger the value selected for $\TD$, the higher the likelihood that transmission on S may be prevented.
By properly selecting this parameter, a compromise can be found between timeliness and spectrum consumption.
As results will show, a significant reduction of latency can be achieved, even for a limited increase of the mean number of transmission attempts per packet performed on air.

Another relevant aspect concerning the above technique is how to select P.
From a general point of view, the channel with the highest QoS should always be chosen, in such a way that the transmission request on S is issued seldom, only when P is experiencing a temporary condition where it suffers from poor communication quality.

\section{Experimental setup}
\label{sec:exp}
\begin{figure}[b]
    \vspace{-1mm}   
    \begin{center}
  	\includegraphics[width=1\columnwidth]{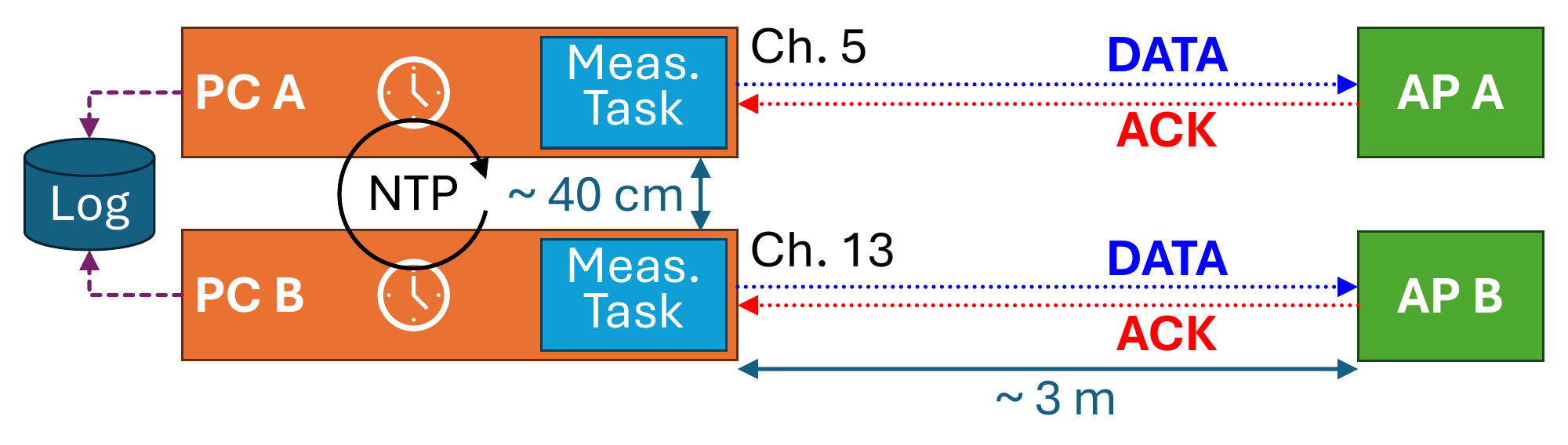}
    \end{center}
    \vspace{-2mm}   
    \caption{Experimental testbed for acquiring the dataset.} 
    \label{fig:testbed}
\end{figure}

The following evaluation is based on experimental data obtained from a real testbed, on which a suitable post-analysis is subsequently carried out.
The testbed we rely on, depicted in Fig.~\ref{fig:testbed}, is similar to what we used in our previous works. 
Nevertheless, the OS has been updated (Ubuntu 24.04.1 LTS, Linux Kernel 6.8.0-45-generic) 
and the measurement application has been improved to log more information for each packet, such as the MCS used for every transmission attempt.
The two PCs, each one provided with a dual-band TP-Link TL-WDN4800 Wi-Fi network adapter,
were set up with the same software.
Their local clocks were synchronized by means of the network time protocol (NTP).
Since PCs are connected to the same floor switch/router through gigabit Ethernet connections, we expect their time to be aligned with an error below $\SI{1}{ms}$.

Every Wi-Fi adapter was set to operate in the $\SI{2.4}{GHz}$ band (quite crowded in our premises, similarly to what happens in many real industrial scenarios, so that non-negligible interference could be experienced without the need to inject artificial traffic) and associated to one of two distinct APs operating on channels $5$ and $13$, respectively.
The two PCs were positioned a few tens of centimeters apart, and the same held for the two APs.
This means that bleeding phenomena related to adjacent channel interference (ACI) were completely prevented.
Instead, the distance between PCs and APs was in the order of three meters.
Since channels $5$ and $13$ are spaced by $\SI{40}{MHz}$, it is unlikely that they may be affected by the same electromagnetic phenomena at the same time (we saw that, near our testbed, channel bonding was used sparingly).

Internal interference caused by the local H/W and S/W (operating system, system buses, caches, etc.)  may sometimes introduce additional delays, unrelated to communication on air, which negatively impact latency.
Unlike \cite{2019-TWC-wired}, having the two links  
managed by distinct (time-synchronized) PCs prevents these phenomena from affecting packets on both links contextually and decreases correlation.

The measurement application on every PC repeatedly performs confirmed point-to-point transmission attempts on its Wi-Fi adapter with a period equal to $\SI{0.5}{s}$.
This time is large enough so that no packet remains queued due to the previous packet still being transmitted.
Since the two PCs are time-synchronized, we may assume that transmission requests are issued approximately at the same time on the two channels.

Information about the transmission of every packet $m_i$ is acquired, including its outcome (either success or failure), the transmission latency $d_i$, the number $a_i$ of transmission attempts actually performed (including retries), and their MCSs.
This was possible by exploiting the software-defined MAC (SDMAC) approach \cite{cena_sdmac_2019}, which requires limited driver modifications by moving the code used to collect measures about communication in user space.
Latency $d_i$ is measured from the transmission request up to the reception of the ACK frame returned by the AP.
From our previous experience, this arrangement provides more precise results than simply forwarding the packet from the AP back to the Ethernet interface of the PC, where a timestamp is taken on its arrival (the latency introduced by our APs is non-negligible and non-constant).
The number of attempts $a_i$ made available by SDMAC permits the evaluation of the actual spectrum consumption.
For the sake of simplicity, we computed it as the mean number $\mu_{a}$ of transmission attempts carried out a packet.

Measurements were collected on the two PCs over a time interval longer than $26$ days, capturing slightly more than $4.5$ million packets on each individual link: A (channel $5$) and B (channel $13$).
Samples in the two logs were first aligned using their timestamps, taken with the synchronized local time of the relevant PC.
Then, exactly $\SI{4500000}{}$ consecutive packet pairs were selected as the dataset for post-analysis.
This is the largest allowed interval for which samples are available on both channels (rounded to the nearest lower boundary of $\SI{100000}{}$ packets).
Since transmission requests for the two packets in every pair were issued approximately at the same time, parallel redundancy can be satisfactorily emulated.
Small deviations of the local clocks are inessential because channels were (practically) independent, and we are interested in an overall statistical evaluation.
To this purpose, the Pearson correlation coefficient was computed for the latency on channels A and B, resulting in $R_\mathrm{AB}=0.086859$.
Although correlation is very small, it is not completely negligible.

\subsection{Characterization of Individual Channels}
The main reason we located the two PCs (which jointly model the STA with redundant links) quite near to the APs is that we wished to emulate a realistic industrial use case where high reliability is demanded.
This affects the deployment of nodes, which should not be positioned too far apart.
By doing so, very few packets were completely lost in the experiment, i.e., seven on channel A and only one on channel~B.
While the reliability of single channels is remarkably good, it is nevertheless much worse than wired networks like (industrial) Ethernet.
Instead, no packets were lost on both channels at the same time, which implies that in our experiment we achieved a PDR equal to $100\%$ for the redundant Wi-Fi link, much closer to wired links.
Clearly, by enlarging the observation interval or by augmenting the distance between the STA and the AP, worse results could be obtained.

\begin{figure}[t]
    \begin{center}
  	\includegraphics[width=1\columnwidth]{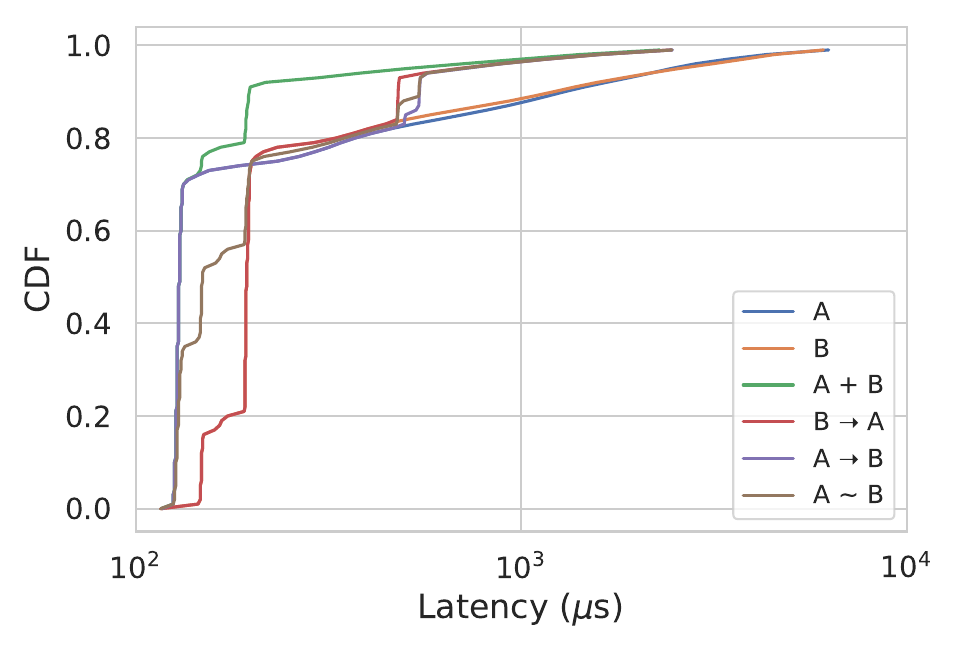}
    \end{center}
    \vspace{-5mm}   
    \caption{CDF of latency for different redundancy approaches.}
    \vspace{-3mm}   
    \label{fig:cdf_latency}
\end{figure}

In our case, the aspect that makes wireless links definitely worse than wired ones is transmission latency.
The best way to characterize this metric is the empirical distribution function, which provides an estimate of the cumulative distribution function (CDF) $F_D(d)$.
It is evaluated as the fraction of correctly delivered packets whose latency is less than or equal to a given value $d$.
Generally speaking, the higher the CDF, the better the QoS.
On a point-to-point cable, and neglecting queuing inside the transmitter and the receiver, the CDF is practically zero up to the frame duration, and then suddenly jumps to a value that is very close to one (some frames may be occasionally lost even on wires, but this is a rare event).
The CDFs on channels A (blue) and B (orange) are depicted in Fig.~\ref{fig:cdf_latency} (a logarithmic scale is used for the X-axis because latency varied in a range wider than three orders of magnitude). 
For low values of $d$, such curves overlap with the violet and red curves, respectively (that will be described later).
As can be seen, there is no clear winner between A and~B: 
on channel A a larger amount of packets experience a low latency (below $\SI{150}{\mu s}$), but channel B becomes slightly better when higher latencies are considered (above $\SI{200}{\mu s}$).

Link behavior can also be summarized by means of conventional statistical indices: mean value ($\mu_d$) and standard deviation ($\sigma_d$), minimum ($d_{\min}$) and maximum ($d_{\max}$), as well as percentiles (which can be found by inverting the CDF).
Statistical indices related to single channels A and B are listed in the first two rows of Table~\ref{tab:stat}.
While the mean latency manages to convey relevant information, the same cannot be said for the standard deviation.
In fact, the shape of the probability density function (PDF) for the latency is very asymmetric.

\begin{table*}
    \caption{Performance indices for redundant approaches: statistics on latency, attempts, and losses ($\TD=\SI{350}{\mu s}$).}
    \vspace{-1mm}   
    \label{tab:stat}
    \centering
    \begin{tabular}{lc|rr|rr|rrrr|cc}
Redundancy mechanism & Symbol & \multicolumn{8}{c|}{Latency ($\SI{}{\mu s}$)} & Mean &   Lost  \\
  & &   $\mu_d$ &  $\sigma_d$ & $d_{\min}$ & $d_{\max}$ & $d_{p50}$ & $d_{p90}$ & $d_{p99}$ & $d_{p99.9}$ & attempts $\mu_{a}$ & packets \\
\hline\hline
only A (non-redundant)    & A &    \SI{533}{} &   \SI{1394}{} &    \SI{117}{} & \SI{215609}{} &    \SI{130}{} &   \SI{1296}{} &   \SI{6270}{} &  \SI{17394}{} & \SI{1.0186}{} &      $7$ \\
only B (non-redundant)    & B &    \SI{566}{} &   \SI{1352}{} &    \SI{116}{} & \SI{443614}{} &    \SI{194}{} &   \SI{1216}{} &   \SI{6074}{} &  \SI{17022}{} & \SI{1.0169}{} &      $1$ \\
both A and B at the same time & A+B  &  \SI{217}{} &    \SI{408}{} &    \SI{116}{} &  \SI{34045}{} &    \SI{130}{} &    \SI{197}{} &   \SI{2285}{} &   \SI{4693}{} & \SI{2.0355}{} &      $0$ \\
\hline
fixed primary A (then B deferred)   & A$\rightarrow$B &   \SI{279}{} &    \SI{451}{} &    \SI{117}{} &  \SI{34045}{} &    \SI{130}{} &    \SI{544}{} &   \SI{2466}{} &   \SI{4915}{} & \SI{1.2304}{} &      $0$ \\
fixed primary B (then A deferred)    & B$\rightarrow$A &  \SI{306}{} &    \SI{431}{} &    \SI{116}{} &  \SI{34395}{} &    \SI{194}{} &    \SI{479}{} &   \SI{2449}{} &   \SI{4831}{} & \SI{1.2171}{} &      $0$ \\
\hline
alternate primary between A and B   &  A$\sim$B     &  \SI{288}{} &    \SI{443}{} &    \SI{116}{} &  \SI{34045}{} &    \SI{149}{} &    \SI{543}{} &   \SI{2459}{} &   \SI{4880}{} & \SI{1.2304}{} &      $0$ \\   
best primary between A and B        & A$^\wedge$B   &  \SI{288}{} &    \SI{439}{} &    \SI{117}{} &  \SI{34045}{} &    \SI{192}{} &    \SI{500}{} &   \SI{2446}{} &   \SI{4842}{} & \SI{1.2080}{} &      $0$ \\
\hline
    \vspace{-4mm}
    \end{tabular}
\end{table*}

The minimum approximates very well the theoretical best case, where a single transmission attempt suffices to correctly deliver the packet and no undue interference is experienced on air, and not even in the PC H/W and S/W.
In our case, $d_{\min}$ includes the durations of the DATA frame we used for channel quality measurement and the related ACK frame, plus one SIFS, and can be computed quite easily if the data rate used for the attempt is known, which in turn depends on the MCS selected by the rate adaptation (RA) algorithm in the driver (Minstrel).

In theory, the maximum is the most important metric in hard real-time systems.
However, it has little practical relevance for two reasons:
a) not necessarily the measured maximum $d_{\max}$ provides a reliable estimate of the worst-case latency that may be theoretically experienced in a real Wi-Fi network, also in those cases (like ours) where the number of considered packets is large; 
b) when using Wi-Fi (or any other wireless communication technologies) for hard real-time systems, the assumption (customarily made in feasibility analysis) that frames are never lost can be hardly justified.
Moving to solutions operating in licensed bands (5G) may improve real-time behavior, but destructive disturbance is practically unavoidable.

Much more useful for conventional industrial distributed control systems are percentiles, defined as the highest latency experienced by a given fraction of the exchanged packets.
For instance, the $99.9\%$ percentile ($d_{p99.9}$) evaluated on a dataset large enough is a satisfactory estimate of the latency value that is exceeded (statistically) only by one packet over a thousand.
Missing a deadline in a control system always has consequences, which in non-critical applications typically result in the quality of the produced good not meeting the intended target.
If this can be tolerated (provided that it happens seldom, as quantified by a specific percentage), the system can be classified as soft real-time, and wireless communication could be used (after link quality has been assessed).
As can be seen in the table, the measured $d_{p99.9}$ for both non-redundant links A and B was found to be in excess of $\SI{17}{ms}$.
If, for example, the deadline for packet delivery is $\SI{5}{ms}$ and the probability that it may be missed must not exceed $0.1\%$, this implies that neither channel A nor B is adequate when used alone.

\subsection{Parallel Redundancy}
By exploiting the very same dataset we used to analyze single channels, the improvements and drawbacks implied by the use of parallel redundancy, denoted A+B for short, were evaluated.
The simplest solution is to issue two contextual transmission requests for every packet, one on every channel.
Since spatial and frequency separation of the two \mbox{Wi-Fi} adapters in our testbed is wide enough, we can reasonably assume that they do not interfere with each other.
Therefore, the performance of parallel redundancy can be determined precisely through post-analysis.
Transmissions in the logs related to channels A and B that have the same timestamp are interpreted as the two redundant copies of the same packet.
Then, the latency experienced when using parallel redundancy is evaluated for every packet $m_i$ as the minimum of the latencies experienced on A and B,
$d_i^{\mathrm{A+B}} = \min(d_i^{\mathrm{A}}, d_i^{\mathrm{B}})$.

Although seamless redundancy is thoroughly covered by previous papers, we decided to include here the results related to our specific dataset, since they characterize the performance in the absence of deferral and are useful for comparison.
Relevant statistics for A+B are reported in the third row of Table~\ref{tab:stat}.
As can be seen, improvements over individual channels
are achieved on every metric: for example,
mean latency $\mu_{d}^{\mathrm{A+B}}$ is less than half with respect to what is experienced on A and B.
However, the main benefits are obtained for high percentiles: $d_{p99}^{\mathrm{A+B}}$ and $d_{p99.9}^{\mathrm{A+B}}$ 
decrease to $\SI{2.3}{ms}$ and $\SI{4.7}{ms}$, respectively, which are about three times lower than  plain \mbox{Wi-Fi}.
When considering the measured maximum $d_{\max}^{\mathrm{A+B}}$, equal to $\SI{34}{ms}$, the improvement is even more pronounced.

However, this comes with a cost: the mean overall number of attempts per packet, 
which quantifies the overall A+B spectrum consumption and can be trivially obtained as the sum of the mean number of attempts performed on A and B, 
$\mu_{a}^{\mathrm{A+B}} = \mu_{a}^{\mathrm{A}} + \mu_{a}^{\mathrm{B}}$,
cannot be less than two.
Often, this cannot be tolerated, because the spectrum is shared among all the STAs of all the wireless networks deployed in the given area, and must be used sparingly.

\section{Transmission Deferral}
\label{sec:def}
A compromise solution that may fit a large number of application scenarios is to proactively decrease the number of transmission attempts performed on air by deferring transmission on one of the two channels.
While A+B is a symmetric solution, we are now considering asymmetric ones, where the primary channel P is involved immediately, whereas the secondary channel S has to wait for the deferral time $\TD$.
If the transmission of packet $m_i$ on P succeeds (as determined by the arrival of the related ACK frame), its transmission request on S is not issued.
In the following, we will denote A$\rightarrow$B the strategy where A is the primary channel and B the secondary channel, and B$\rightarrow$A the case where roles are inverted.

Results are reported in the fourth and fifth rows of Table~\ref{tab:stat} for $\TD = \SI{350}{\mu s}$
($\sim3\cdot d_{\min}$, a satisfactory compromise that makes $d_{p99.9} < \SI{5}{ms}$).
As can be seen, the mean latency value is higher than the A+B case, but remains well below individual channels A and B.
Concerning high percentiles, which are the most relevant performance indices in control systems, they are only slightly worse than the A+B case, and the same holds for the measured maximum.
On the other hand, concerning spectrum consumption, we see that the mean number of transmission attempts on air is about $22\%$ higher than single channels, and much lower than parallel redundancy.
This means that deferral techniques provide a good trade-off between dependability/timeliness and resource consumption.

Curiously, A$\rightarrow$B provided better results than B$\rightarrow$A for the mean latency and the distribution median (that coincides with $d_{p50}$), but is slightly worse when high percentiles are taken into account.
This depends on the particular shape of the latency CDFs on A and B, according to which no channel is definitively better than the other.

\subsection{Choosing the Deferral Time}
By properly selecting $T_\mathrm{D}$, the number of transmission attempts on S can be lowered at will.
Generally speaking, when $T_\mathrm{D}$ is increased a larger fraction of packets will be transmitted only on P.
However, this also makes latency on the redundant link higher, because we are purposely delaying channel S.
This means that some compromise should be found.
To quantitatively assess the effects of $T_\mathrm{D}$ we carried out a post-analysis where performance metrics were evaluated for different values of this parameter (in $\SI{}{\mu s}$), chosen in the set $\{150,\!250,\!350,\!450,\!550,\!650,\!750,\!850,\!950,\!1150,\!1350,\!1550\}$, which spreads over one decade.
Applying deferral during post-analysis is acceptable, since what really matters is the statistical characterization of channels given by experimental logs, which does not change for small time displacements.

\begin{figure}[b]
    \vspace{-5mm}
    \begin{center}
 	\includegraphics[width=0.95\columnwidth]{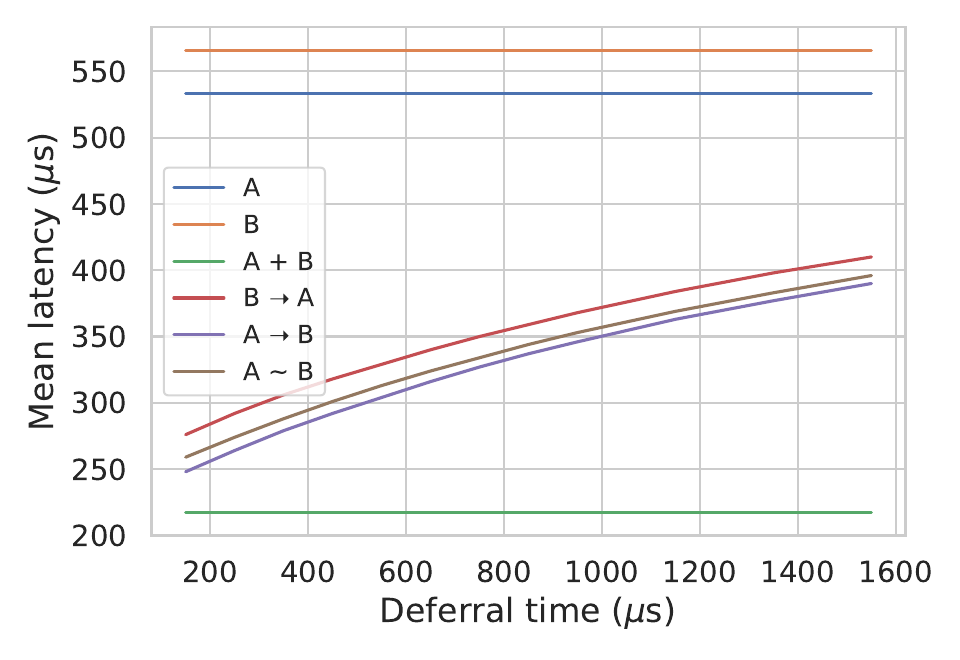}
    \end{center}
    \vspace{-5mm}
    \caption{Mean latency ($\mu_d$)}
    \label{fig:mean}
\end{figure}

\begin{figure}[b]
    \vspace{-5mm}   
    \begin{center}
 	\includegraphics[width=.95\columnwidth]{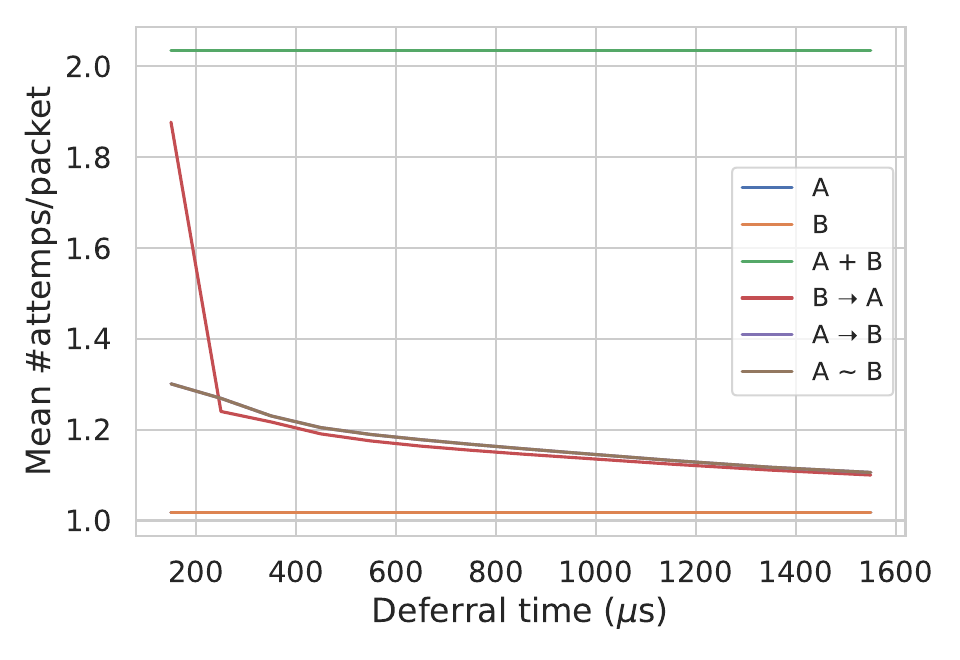}
    \end{center}
    \vspace{-5mm}
    \caption{Mean number of attempts per packet ($\mu_{a}$).} 
    \label{fig:eff}
\end{figure}

Figs.~\ref{fig:mean} and~\ref{fig:eff} show the mean latency and mean number of attempts per packet, respectively.
The blue, orange, and green curves refer to modes A, B, and A+B, which do not depend on $T_\mathrm{D}$ and hence are constant.
Concerning spectrum consumption in Fig.~\ref{fig:eff}, curves for single channels A and B (at the bottom) overlap, and the same holds for curves describing  \AtB (violet) and \alt (brown, will be described later).


Concerning any single packet $m_i$, parallel redundancy cannot exhibit a behavior worse than any of the constituent channels.
Therefore, in the absence of losses (or when they are negligible), mode A+B  always outperforms both A and B, whatever statistical index is considered (mean, percentile, etc.).
Likewise, deferred redundant modes always behave better than their primary channel P considered alone.
As a consequence, A$\rightarrow$B is always better than A and B$\rightarrow$A is always better than B.
For these reasons, lines A and B in Fig.~\ref{fig:mean} can be reasonably seen as upper bounds for the mean latency of all redundant modes, whereas A+B constitutes a proper lower bound.

Figures confirm our reasoning: when $T_\mathrm{D}$ is increased, mean latency progressively grows, while spectrum consumption decreases.
In the latter case, however, a sudden decrease may be initially observed, after which $\mu_{a}$ improves slowly.
At any rate, when $T_\mathrm{D}$ reaches $\SI{1.55}{ms}$, spectrum consumption is just $10\%$ higher than for single non-redundant links, which is a relevant result.

\begin{figure}[t]
    \begin{center}
 	\includegraphics[width=0.95\columnwidth]{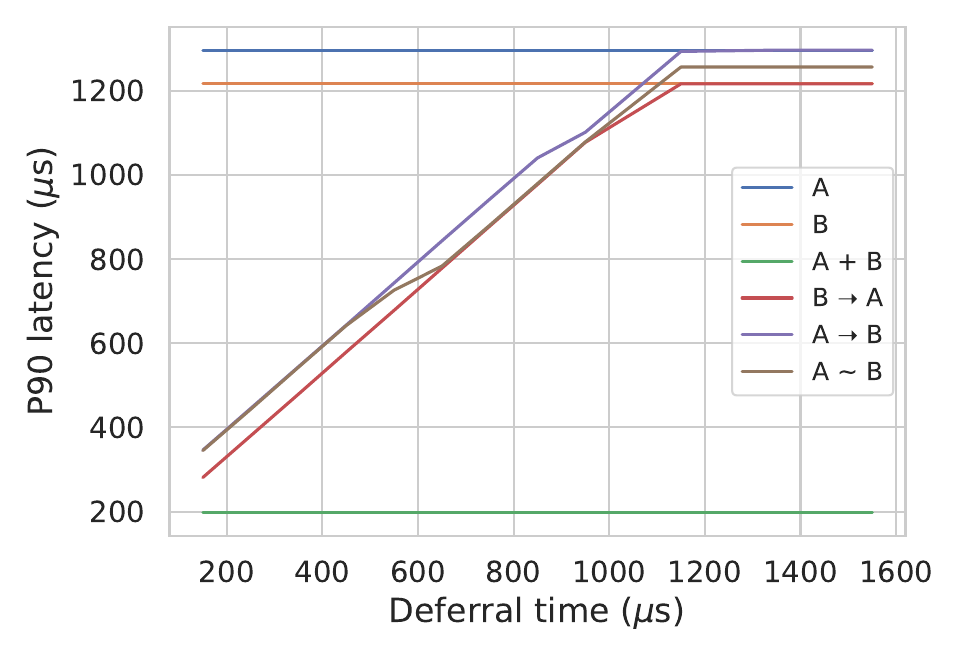}
    \end{center}
    \vspace{-5mm}
    \caption{90-percentile of latency ($d_{p90}$).}
    \vspace{-5mm}
    \label{fig:P90}
\end{figure}

\begin{figure}
    \begin{center}
 	\includegraphics[width=0.95\columnwidth]{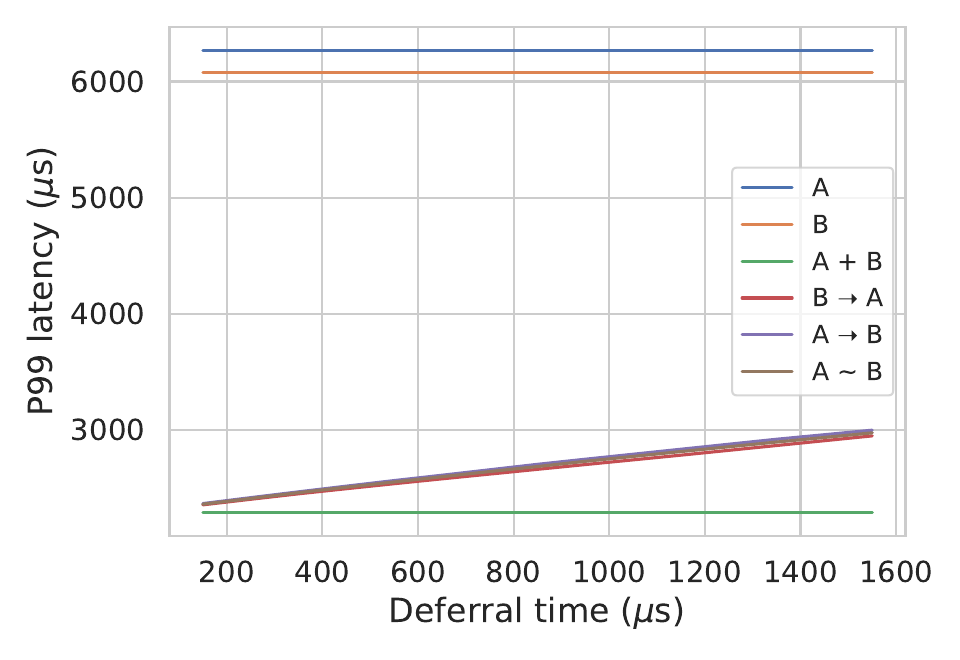}
    \end{center}
    \vspace{-5mm}
    \caption{99-percentile of latency ($d_{p99}$).}
    \vspace{-5mm}
    \label{fig:P99}
\end{figure}

\begin{figure}
    \vspace{-5mm}
    \begin{center}
 	\includegraphics[width=0.95\columnwidth]{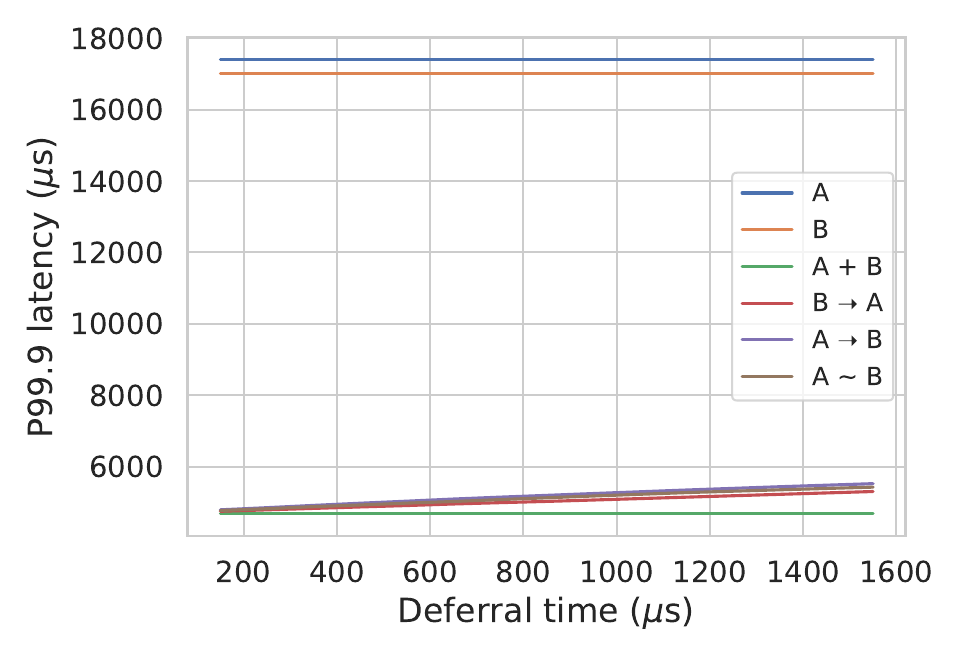}
    \end{center}
    \vspace{-5mm}
    \caption{99.9-percentile of latency ($d_{p99.9}$).}
    \label{fig:P99.9}
\end{figure}

To better appreciate this, three high percentiles on latency are shown in Figs.~\ref{fig:P90}, \ref{fig:P99}, and~\ref{fig:P99.9} ($d_{p90}$, $d_{p99}$, and $d_{p99.9}$, respectively) versus the deferral time.
As can be seen, percentiles for redundant modes worsen as $\TD$ increases, which is expected since we are delaying the secondary channel.
The shape of curves depends on the CDF of individual channels A and B, and the related model is expected to be quite complex.
However, the growing trend of percentiles versus $\TD$ appears to be approximately linear.
Fig.~\ref{fig:P90} agrees with our previous explanations about upper bounds.
In particular, $d_{p90}$ for redundant deferred modes increases in an approximately linear way until it reaches $d_{p90}$ of the related primary channel, which cannot be exceeded.
The very same behavior is observed in Figs.~\ref{fig:P99} and \ref{fig:P99.9} for $d_{p99}$ and $d_{p99.9}$, with the only difference that growth is slower.
Until the point where plots converge, deferral is advantageous for soft real-time systems.

Concerning the position of that point, the following reasoning can be done.
Let us consider, for example, the dataset related to A. 
Percentile $d_{p90}^\mathrm{A}$ is obtained as follows: 
first, the set of packets is sorted in increasing latency order; 
then, a subset is obtained by retaining $90\%$ of the initial set; 
and finally, the latency of the last packet in the subset (the highest one) is taken as $d_{p90}^\mathrm{A}$
(since our dataset was quite large, skipping interpolation leads to a negligible error).
Now, if deferred redundancy \AtB  is exploited, 
no benefits can be achieved concerning $d_{p90}^{\mathrm{A}\rightarrow\mathrm{B}}$
when $d_{\min}^\mathrm{B}+\TD > d_{p90}^\mathrm{A}$, because: 
a) none of the packets that suffer from a latency strictly higher than $d_{p90}^\mathrm{A}$ can take advantage of transmission on B to lower their latency below $d_{p90}^\mathrm{A}$; and, 
b) nothing changes for packets whose latency is strictly lower than $d_{p90}^\mathrm{A}$.
Hence, $\TD^* = d_{p90}^\mathrm{A} - d_{\min}^\mathrm{B}$ 
represents the point where $d_{p90}^{\mathrm{A}\rightarrow\mathrm{B}}$ saturates and becomes equal to $d_{p90}^\mathrm{A}$.
A similar reasoning also holds for the other percentiles 
and when comparing B and B$\rightarrow$A.

For $d_{p99.9}^\mathrm{A}$ and $d_{p99.9}^\mathrm{B}$, $\TD^*$ is located near $\SI{17}{ms}$,
which explains the low values (with respect to individual channels)
we observed 
for $d_{p99.9}^{\mathrm{A}\rightarrow\mathrm{B}}$ and $d_{p99.9}^{\mathrm{B}\rightarrow\mathrm{A}}$
in the entire range $\TD \in [\SI{150}{\mu s},\SI{1550}{\mu s}]$.
When $\TD$ is set to $\SI{1.55}{ms}$, $d_{p99.9}$ remains below
$\sim\SI{5.5}{ms}$ for deferred redundancy modes, with a cost, in terms of the increase in the mean number of attempts performed on air, which remains near $10\%$.
This is much better than $d_{p99.9}$ for non-redundant links, which is above $\SI{17}{ms}$,
and only slightly worse than $d_{p99.9}^\mathrm{A+B} = \SI{4.7}{ms}$ for parallel redundancy A+B, which consumes much more spectrum.
Hence, deferred redundancy is worth being employed.


\section{Improving Deferred Redundancy}
\label{sec:alt}
The main problem of static solutions like \AtB and \BtA is that, if the worst channel is accidentally selected as primary, redundancy provides suboptimal performance.
This effect is only partially visible with our dataset because, unlike \cite{2019-TWC-wired}, the behavior of the two channels was pretty similar.
Nevertheless, the presence in the real world of temporary phenomena like interference (several STAs tuned on the same channel that start exchanging high volumes of data), path loss and multipath fading (especially when the STA is moving), and narrowband electromagnetic noise (that can be emitted by high-power industrial equipment), may result in rather different behaviors between the channels that make up the redundant link.
In these cases, a bad selection of P has a significant impact on communication quality, and hence on the operation of distributed applications.

To get the most out of deferred redundant approaches, the best channel should be identified on a per-packet basis and used as the primary one, which leads us into the realm of channel quality prediction for Wi-Fi \cite{2022-ITL-ML}.
In this paper, two basic solutions were analyzed that do not employ machine learning (ML).
The former is a very simple one, in which the roles of P and S are alternated on every packet, whereas the second relies on more refined heuristics.

\subsection{Alternating Deferred Redundancy}
This redundancy mode alternatively behaves as \AtB and B$\rightarrow$A.
No attempts are made to determine the best channel, which means that, on average, half of the choices are good and half are bad.
What we expect from this arrangement, we denote with the notation A$\sim$B, is that the QoS on the redundant link stays in between \AtB and B$\rightarrow$A, and constantly remains much better than single channels, irrespective of what happens to the disturbance on them.

Results, obtained by setting $\TD = \SI{350}{\mu s}$, are reported in the sixth row of Table~\ref{tab:stat}.
As can be seen, behavior actually resembles what we expect, which is good news because both \AtB and \BtA are advantageous.
This is confirmed by looking at Figs.~\ref{fig:mean}--\ref{fig:P99.9}, where the curves referring to the alternating deferred redundancy A$\sim$B (colored in brown) are always found between A$\rightarrow$B (violet) and
B$\rightarrow$A (red).

\subsection{Heuristics-based Deferred Redundancy}
The alternating \alt approach is quite simple and is acceptable when the quality of the two constituent channels is more or less the same.
However, when this is not true (for instance, when one of the channels is set in the overcrowded $\SI{2.4}{GHz}$ band while the other operates in the $\SI{5}{GHz}$ band, where very old and cheap devices are absent and higher data rates are possible), this strategy is suboptimal.

To overcome these limitations, we defined simple heuristics to determine, for every single packet, which channel is expected to provide the best choice.
The solution we devised for this purpose, we denote A$^\wedge$B, is quite basic and combines mechanisms to deal with the following aspects:
\begin{itemize}
    \item 
    Locality of issues in the \textit{short term}: network/spectrum phenomena that cause a packet to suffer from high latency may last for some time before ending. 
    Therefore, they may affect the following transmissions on the same channel.
    In our case, packets are spaced quite wide ($\SI{0.5}{s}$) and hence we should expect very low autocorrelation.
    We denote the case of S being faster than P for packet $m_i$ (that is, not only $d_i^\mathrm{P} > \TD$, which causes transmission on S as well, 
    but also $d_i^\mathrm{S} + \TD < d_i^\mathrm{P}$) as a \textit{bad guess} event.  
    \item
    Locality of issues in the \textit{mid term}: the average quality of every channel is ever changing, and the same applies to the channel deemed to be the \textit{best} at any given time.
    To determine the optimal choice (i.e., the channel C with the lowest mean latency), an exponential moving average (EMA) can be exploited with a smoothing factor $\alpha$.
    In this case, the estimate $\hat{d}_i^\mathrm{C}$ is evaluated as 
    \begin{align}
        \label{eq:dhat}
        \hat{d_i^\mathrm{C}} = \alpha \cdot d_i^\mathrm{C} + (1-\alpha) \cdot \hat{d}_{i-1}^\mathrm{C}
    \end{align}
    \item
    No transmission request is made on S when $d_i^\mathrm{P} < \TD$, and hence its quality can not be evaluated through Eq.~\ref{eq:dhat}.
    The lack of updates about latency, losses, and the number of attempts performed on S worsens future predictions for it.
    Occasionally selecting the worst channel (exploration) instead of the best one (exploitation) may help.
\end{itemize}

The way the above mechanisms were combined in our heuristics is quite trivial: we mixed them by changing the probability with which they were selected, by also varying $\alpha$, and performed several trials by hand to determine what configuration offers the best results.
Unfortunately, the behavior of real channels acquired in the experimental campaign did not differ significantly and did not vary consistently over the course of the experiment.
For this reason, only limited benefits were obtained by our heuristic-driven deferred redundancy.

Results for \best are reported in the seventh (and last) row of Table~\ref{tab:stat}.
As can be seen, both high percentiles and spectrum consumption decreased slightly with respect to A$\sim$B,
which is anyway valuable for soft real-time systems.

\section{Conclusions}
\label{sec:conc}
While wireless communication, and in particular Wi-Fi, are increasingly adopted in industrial environments to enable moving devices to communicate with each other and with the factory backbone, they are still deemed not dependable enough for exchanging process data in real-time control loops.
As pointed out several times in the literature, seamless redundancy may tangibly increase dependability by decreasing both losses and latency.
However, it also implies higher spectrum consumption, which sometimes cannot be tolerated.

In this paper, the concept of deferred parallel redundancy, introduced a few years ago, is thoroughly investigated.
Specifically, transmission requests on the secondary channel are temporarily deferred and only take place if, in the meantime, no confirmation has arrived on the primary channel.
As part of this work, this approach has been improved with techniques aimed at optimal primary channel selection at runtime, and its advantages and drawbacks in terms of latency, packet losses, and spectrum consumption were analyzed using several metrics.

Results, obtained through post-analysis carried out on a large experimental dataset we purposely acquired from scratch, show that tangible advantages can be practically achieved by deferred redundancy for high latency percentiles, which are the most relevant metrics in soft real-time control systems.
In particular, they can be made about three times smaller than non-redundant links, by consuming just $10\%$ more spectrum resources.

With such techniques, the ability to predict with reasonable certainty what is the best choice for the primary channel on a per-packet basis becomes essential to achieve optimal results.
Simple heuristics were devised for this purpose, which unfortunately provided limited benefits with our dataset.
We strongly believe that noticeably better results could be obtained by exploiting machine learning for channel selection, and this will be part of our future work.
Besides, we'll try to apply deferred redundancy to MLDs, also considering frame aggregation.

\bibliographystyle{IEEEtran}
\bibliography{bibliography}

\end{document}